\documentclass[twocolumn,aps]{revtex4}
\usepackage{graphicx}% Include figure files
\usepackage{dcolumn}% Align table columns on decimal point
\usepackage{bm}% bold math\setlength{\textwidth}{17cm}
\begin{document}

\title{Effects of diffusion and particle size in a
kinetic model of catalyzed reactions}

\author{T.G. Mattos}
 \email{tgmattos@if.uff.br}
 \affiliation{Instituto de F\'{\i}sica, Universidade Federal Fluminense, 
Av. Litoranea s/n, Campus da Praia Vermelha, Niteroi RJ 24210- 340, Brazil.}

\author{F\'abio D. A. Aar\~ao Reis}
 \email{reis@if.uff.br}
 \altaffiliation[~~Also at ]{Department of Chemistry, University of Wisconsin - Madison, WI 53706, USA.}
 \affiliation{Instituto de F\'{\i}sica, Universidade Federal Fluminense, 
Av. Litoranea s/n, Campus da Praia Vermelha, Niteroi RJ 24210- 340, Brazil.}

\date{\today}% It is always \today, today,

\begin{abstract}
We study a model for unimolecular reaction in a supported catalyst
including reactant diffusion and desorption, using analytical methods
and scaling concepts. For rapid reactions, enhancing surface diffusion or
increasing particle size favors the flux of reactants to the catalyst
particles, which increases the turnover frequency (TOF). The reactant flux
towards the support becomes dominant when the ratio of diffusion lengths in
the catalyst and in the support exceeds a critical value. A peak in the TOF
is obtained for temperature-dependent rates if desorption energy in the
support ($E_d$) exceeds those of diffusion ($E_D$) and reaction ($E_r$).
Significant dependence on particle size is observed when
the gaps between those energies are small, with small particles giving higher
TOF. Slow reactions ($E_r>E_d$) give TOF monotonically increasing with
temperature, with higher reactant losses in small particles. The scaling concepts
can be extended to interpret experimental data and results of more complex models.
\end{abstract}

%Use showkeys class option if keyword
\maketitle

\section{Introduction}

Modeling heterogeneous catalytic processes is an essential
tool for catalyst design and for the improvement of operating conditions
\cite{broadbelt2000,zhdanovrev,lynggaard,murzin}.
Hierarchical approaches have to be adopted due to the need of information on a
wide range of length and time scales, from the electronic structure to
the reactor design \cite{broadbelt2000}. An important step of this approach is
the microkinetic modeling, where microscopic processes such as reaction,
diffusion, aggregation, and desorption are described by stochastic rules,
providing information on the efficiency of the catalytic process in
length scales ranging from a few nanometers to several micrometers.

% trocar termo spillover, mas comentar sobre a diferenca
An important problem to be addressed with these methods is the effect of
diffusion of reactants through the interface between the catalyst particle and
the support. Several recent experimental papers illustrate these phenomena in
catalyzed reactions
\cite{piccolo,chen,polychrono,kecskemeti,laurin,odier,marques,bowker,mao}
as well as in related problems, such as gas adsorption \cite{lueking,dutta,jain},
where the same materials may be used.
Morphological features of the catalyst and support and
physicochemical conditions of operation determine the role of diffusion
on the performance of the process \cite{zhdanovrev,henryrev}.
Sometimes these phenomena are called capture-zone effects, since a
certain region of the support surrounding the particle may increase the
effective area for capturing reactants from the gas phase, but there are
also cases where a net flux to the support is observed. When only one phase
can adsorb the reactant from the gas, the terms
spillover (reactant flux from the catalyst to the support) or back
spillover (the opposite flow) are used \cite{conner,libuda},
although recently many authors have extended these terms to reactants
adsorbing in both phases \cite{zhdanovrev} (and here we will use them to
facilitate the discussion of the results).

% citar trabalho mais recente do Cwiklik
Due to the large interest in industry, some models which incorporated
the effects of reactant diffusion through the catalyst-support interface were
designed for certain applications.
The simplest models are based on rate equations (mean-field models) that do
not account for the spatial heterogeneity of the media where reactions take
place \cite{hoffman,costa,christou,galdikas}, or that use some type of
approximation to represent that heterogeneity \cite{dooling1999,cwiklikCPL}.
Other models represent it through distributions of catalytic sites in
lattices. Most of them are designed to describe $CO$ oxidation in different
catalysts and conditions \cite{henryJCP,mcleod,zhdanov1997,johansson}, and
simplify diffusion and adsorption processes of some species (although
other applications have also been proposed
\cite{jain,mcleod,mcleodCERD2004,mcleodCES2004}).
In order to get a deeper insight on the effects of reactant diffusion,
Cwiklik et al \cite{cwiklikSS,cwiklikASS} recently simulated simple
reaction-diffusion models in surfaces with catalytic stripes and squares, as
well as random distributions of catalytic sites. In certain ranges of
parameters, they showed monotonic dependences of the turnover frequency (TOF)
on diffusion coefficients and reaction rates \cite{cwiklikSS}.

A number of other papers aim at a full investigation of simple models of
reaction and diffusion. In the present context, relevant examples are models in
lattices with distributions of catalytic sites, i. e. with some type of
non-homogeneity of catalytic activity
\cite{oshaninJCP,benichou,oshaninPRL,albano1,albano2,jansenPRL}.
Even adsorption-desorption models without surface diffusion show
that the correlations in catalyst particle position have nontrivial
effects on the TOF, independently of adsorbate interactions
\cite{oshaninPRL,rieger,oshaninJSP}. Moreover, models including diffusion in
heterogeneous
surfaces show that the structures that maximize the efficiency of a
catalytic process are highly dependent on the rates of the main
microscopic processes \cite{jansenPRL}. 

% explicar que o nosso modelo e' um lattice model (ref 1.1)
% particular applications changed (ed 2)
In the present paper, we will propose a one-dimensional lattice model for
unimolecular reactions in a supported catalyst, with reactant diffusion and
desorption both in the support and in the catalyst particles. Our aim is to
understand the interplay between these physicochemical mechanisms, particle
size and catalyst coverage. The model geometry is equivalent to that of Ref.
\protect\cite{cwiklikSS}, but here we will obtain an analytic solution that
facilitates the illustration of different possible outcomes. We will use
scaling concepts to explain the model results, so that this framework can
be extended to more complex models and applications to real systems, where
numerical solutions are usually necessary. 
Indeed, scaling approaches were already shown to be very useful
to understand qualitative trends
in reaction-diffusion models \cite{incubation}.

% retirada referencias a Cwiklik na 1a frase (ref. 1.8)
Among our results, we will distinguish conditions to enhance the net reactant
flux from the catalyst to the support or vice-versa by varying one of the
microscopic rates, and we will discuss the effect of increasing the
catalyst particle size. The increase or decrease of the TOF will be shown
to depend on the relation between diffusion lengths of reactants in the
catalyst and in the support. Some of these results reinforce findings
of previous works \cite{mcleod,zhdanov1997,cwiklikSS}. Moreover, under
reasonable assumptions for temperature-dependent rates, we will show that a
remarkable increase in the TOF can be obtained if reverse
spillover regularly fills the catalytic particles with the reactants adsorbed
in the support. This feature may be observed in a large temperature range, a
possibility which is interesting for applications.
The identification of these scenarios is possible because the model
accounts for the inhomogeneity of the catalytic system and, consequently,
predicts inhomogeneous distributions of reactants, which advances over the
mean-field models.

% paragrafo "The rest of this work..." removido (ref. 2.2)
% esta remocao segue o padrao dos artigos do J. Cat. (ver refs. 21, 22, 28,
% 50)

% title changed (ref. 2.3)
\section{The model}

The physicochemical processes involved in the model are illustrated in Fig. 1.

\begin{figure}[!h]
\includegraphics[width=0.5\textwidth]{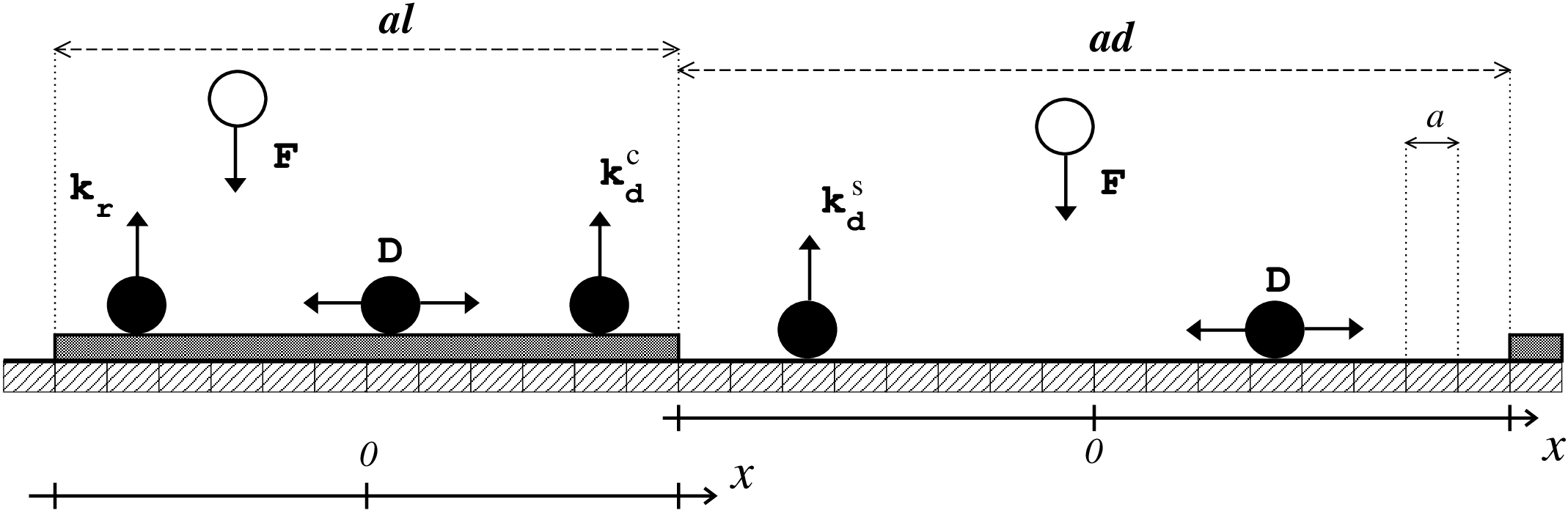}
\caption{\footnotesize Schematic representation of the model for the supported catalyst, with the rates of the physicochemical processes. Non-adsorbed reactant A (equation \ref{reactionER}) is represented by white circles, while adsorbed reactant A is represented by black circles. The $x$ axis used in the analytic solution are shown (for the particle and for the support).}
\label{fig1}
\end{figure}

% removed the term "adsorption" to define catalytic sites because all sites
% can adsorb (ref. 1.9)
The catalyst particles are represented by segments of $l$
sites in a line, separated by $d$ support sites. Assuming that $a$ is the size
of the lattice site, this corresponds to particles of diameter $la$
separated by a distance $da$. The fraction of
the support covered by the catalytic material is 
\begin{equation}
\epsilon \equiv \frac{l}{l+d} .
\label{defrho}
\end{equation}
This lattice structure may be a reasonable description of some model catalysts
\cite{libuda,somorjai}, as
well as a good approximation to the morphology of catalytic clusters supported
in materials with long and narrow pores \cite{schuth}.

% comentarios sobre valor de "a" para justificar valores de "l" (ref 1.3)
The particle size is determined by the physico-chemical
conditions in which the catalytic material is deposited on the support
\cite{freund2002}. The fraction of the support covered, $\epsilon$, is related
to the amount of material used to produce a sample as well as the particle
shape,
which determines the surface to volume ratio.
In oxide supported $Pt$ or $Pd$ catalysts, the particle diameter usually
vary from $1$ to $50~ nm$ - depending on the growth conditions and if sintering
occurs, sizes of $100~ nm$ or more can be found. Since $a$ is of order of a few
angstroms, this typically corresponds to $l$
ranging from $3$ to $150$ lattice sites. The spacing between the particles has a
broad distribution in catalysts supported in porous materials, but in model
catalysts they are nearly uniform, usually in the range $50-200~ nm$
($d$ typically between $100$ and $500$ lattice spacings for most oxide supports).

The flux of a single chemical species (reactant) towards the surface occurs with
rate $F$, which is defined as the number of incoming molecules per site
per unit time. In most of this work, $F$ will be used to define the time scale
of
the model, so that other frequencies are calculated relatively to this quantity.
The reactant adsorbs in the (randomly chosen) site of incidence
if it is empty, either in the catalyst or in the support, otherwise the
adsorption
attempt is rejected. Sticking coefficients are set equal to $1$, since the
effects
of different values in the catalyst and in the support can be incorporated in
the
corresponding desorption rates: $k_d^s$ and $k_d^c$, in the support and in the
catalyst
particles, respectively (each rate corresponds to number of events per site per
unit
time). Interaction between the adsorbates is restricted to the excluded volume
condition.

Adsorbed reactants diffuse with coefficient $D$, which for simplicity is assumed
to be the same in the support and in the catalyst. It means that each reactant
attempts to execute $2D/a^2$ random steps to nearest neighbor sites per unit
time. We are assuming that the activation energy for diffusion is the same in
those regions, which is certainly not true for a real catalyst. However, it is
not a very restrictive assumption for our study because the results are
interpreted in terms of diffusion lengths and scaling
ideas are emphasized, allowing a direct extension to cases of different $D$ in
different regions.

% explicacao sobre mecanismo de difusao (ref 1.4)
Since reactants are always in contact with the support or the catalyst, this
is a surface diffusion model, which is a reasonable assumption on metal particles.
On the other hand, Knudsen diffusion may be more realistic for the movement of
reactants inside a catalyst pore. However, this does not represent a limitation 
because a suitable value of $D$ can be chosen, and interpretations based
on diffusion lengths are still valid.

% species C arriving FROM the gas phase - correcoes em dois pontos
% (ref. 1.10 e 1.11)
The adopted reaction mechanism for reactant $A$ and product $B$ is
\begin{equation}
A_{ads} \rightarrow B_{gas} ,
\label{reaction}
\end{equation}
with reaction rate $k_r$ assumed to be uniform in the catalytic
region (this rate corresponds to number of events per site per unit time).
The unimolecular reaction (\ref{reaction}) may represent an Eley-Rideal (ER)
mechanism, in which the adsorbed species reacts with another species $C$ arriving
from the gas phase and forms a volatile product:
\begin{equation}
A_{ads} + C_{gas} \rightarrow B_{gas} .
\label{reactionER}
\end{equation}
For instance, an application to $CO$ oxidation is discussed in Refs.
\protect\cite{costa,christou}. In this case, $k_r$ not only accounts for
activation of the adsorbed reactant but also for the flux of the other reactant
from the gas phase. Certainly the assumption that $k_r$ is constant in the whole
catalytic region is not realistic because it is well known that different
crystalline faces of a metal have distinct catalytic activity. However, the
present assumption is useful for a study which aims at investigating the
interplay of many other different physico-chemical processes.

% title changed
\section{Analytic solution}

In order to solve the model analytically, we assume that the catalyst particles
and the support segments between them are sufficiently large ($l\gg 1$, $d\gg
1$), so that a continuous approximation is possible. Although $l$ and $d$ may
not be very large in real systems, we will show that the continuous
approximation
works well even with $l$ and $d$ of order $10$.

% indicar catalyst e support antes das equacoes (ed 3)
% explicar os dominios de x no catalisador e no suporte (ref 5)
The dimensionless reactant coverages in the catalyst particle and in the support
are respectively defined as $\theta_{\rm c}\left( x,t\right)$ and
$\theta_{\rm s}\left( x,t\right)$.
For simplicity, we use the same variable $x$ for position in both regions, with
the range $-la/2\leq x\leq la/2$ in the catalyst and the range
$-da/2\leq x\leq da/2$ in the support.
Diffusion, reaction and adsorption-desorption processes of Fig. 1 lead to
equations for surface coverages; in the catalyst, we have
\begin{equation}
\frac{\partial}{\partial t} \theta_{\rm c}(x,t) = D
\frac{\partial^2}{\partial x^2} \theta_{\rm c}(x,t) + F
\left[ 1-\theta_{\rm c}\left( x,t\right)\right] - ( k_{\rm r} +
k_{\rm d}^{\rm c} ) \theta_{\rm c}(x,t) ,
\label{eqcat}
\end{equation}
and in the support we have
\begin{equation}
\frac{\partial}{\partial t} \theta_{\rm s}(x,t) = D
\frac{\partial^2}{\partial x^2} \theta_{\rm s}\left( x,t\right) +
F \left[ 1-\theta_{\rm s}\left( x,t\right)\right] -
k_{\rm d}^{\rm s} \theta_{\rm s}\left( x,t\right) .
\label{eqsup}
\end{equation}
Each of these equations is valid in the above defined ranges of $x$.

Here we are interested in steady state solutions, where 
$\frac{\partial}{\partial t} \theta_{\rm c}(x,t) = \frac{\partial}{\partial t}
\theta_{\rm s}(x,t) = 0$ and, consequently,  $\theta_{\rm c}$ and $\theta_{\rm
s}$ depend only on $x$. In the catalyst, this gives
\begin{equation}
D \frac{\rm{d}^2 \theta_{\rm c}}{\rm{dx}^2} + 
F\left( 1-\theta_{\rm c}\right)  -  {\left( k_{\rm r} + k_{\rm d}^{\rm c}
\right)}
\theta_{\rm c} = 0 ~ .
\label{eqcatss}
\end{equation}
In the support, an analogous equation (without the
reaction term) is obtained. Eq. (\ref{eqcatss}) can be easily solved and gives
\begin{equation}
\theta_{\rm c}(x) = r_{\rm c} + \alpha_{\rm c}\cosh{(x/\lambda_{\rm c})} ,
\label{solthetacat}
\end{equation}
where $\alpha_c$ is a constant to be determined from boundary conditions and
\begin{equation}
r_{\rm c} \equiv  \frac{1}{1 + k_{\rm r}/F + k_{\rm d}^{\rm c}/F} \qquad ,
\qquad
\lambda_{\rm c} ~\equiv~ \sqrt{\frac{D/F}{1 + k_{\rm r}/F + k_{\rm
d}^{\rm c}/F }} .
\label{defrclambdac}
\end{equation}
Analogously, in the support we obtain
\begin{equation}
\theta_{\rm s}(x) = r_{\rm s} + \alpha_{\rm s}\cosh{(x/\lambda_{\rm s})} ,
\label{solthetasup}
\end{equation}
where $\alpha_s$ is a constant and
\begin{equation}
r_{\rm s} \equiv \frac{1}{1 + k_{\rm d}^{\rm s}/F} \qquad , \qquad
\lambda_{\rm s} \equiv \sqrt{\frac{D/F}{1 + k_{\rm d}^{\rm s}/F }} .
\label{defrslambdas}
\end{equation}

Note that, as expected, diffusion, reaction and desorption rates appear in our
results in the form of ratios to the external particle flux $F$.

The calculation of unknown constants in equations such as (\ref{solthetacat})
and
(\ref{solthetasup}) usually follows from the use of suitable boundary
conditions.
However, here diffusion leads to a net flux of reactants from the catalyst to
the support or vice-versa, which depends on the
competition of all other processes along both regions. Thus, those unknown
constants
will be determined by matching the gain and loss terms in each region due to all
those processes, and the net flux at the interfaces will be obtained from them.

% explicitar valor de x nas eqs. 12 e 13 (ref. 1.12)
% trocar last site por edge site (ed 4)
In the catalyst, the contribution to the loss rate due to diffusion involves the
probability of finding a reactant in the edge site of that region and of finding
the neighboring support site empty (due to the excluded volume condition).
Other contributions come from reaction and desorption along the particle.
Thus, the loss in the coverage of the catalytic region per unit time is
\begin{eqnarray}
\left( \Delta\theta_{\rm c} \right)_{loss} & = & \frac{2D}{a^2} \left( 1 -
\theta^{*} \right) \theta^{\dagger} + \frac{1}{a}\int_{-la/2}^{la/2}{ ( k_{\rm
r} + k_{\rm
d}^{\rm c} )~\theta_{\rm c}(x')~ dx'} \nonumber \\
& = & \frac{2D}{a^2} \left( 1 - \theta^{*} \right)\theta^{\dagger} +
l\left( k_{\rm r} + k_{\rm d}^{\rm c} \right) \bar{\theta_{\rm c}} ,
\label{dtcloss}
\end{eqnarray}
where $\theta^{*}$ is the coverage of the edge site of the support region
(neighbor of the catalyst)
\begin{equation}
\theta^{*} \equiv \theta_{\rm s}\left( x=\frac{da}{2}\right) = r_{\rm s} +
\alpha_{\rm s}\cosh{ \left( \frac{da}{2\lambda_{\rm s} } \right) } ,
\label{thetaast}
\end{equation}
$\theta^{\dagger}$ is the coverage of the edge site of the catalytic region
(neighbor of the support)
\begin{equation}
\theta^{\dagger} \equiv \theta_{\rm c}\left( x=\frac{la}{2}\right) = r_{\rm c} +
\alpha_{\rm c}\cosh{ \left( \frac{la}{2\lambda_{\rm c} } \right) } ,
\label{thetadag}
\end{equation}
and $\bar{\theta_{\rm c}}$ is the average coverage of the catalytic region
\begin{equation}
\bar{\theta_{\rm c}} \equiv \frac{1}{la}\int_{-la/2}^{la/2}{ \theta_{\rm c}(x')
dx' } .
\label{thetacmedio}
\end{equation}
The gain rate, which accounts for flux from the support to the catalyst
at the edge sites and for the external flux, is
\begin{eqnarray}
\left( \Delta\theta_{\rm c} \right)_{gain} & = &
\frac{2D}{a^2} \theta^{*}( 1 - \theta^{\dagger}  ) +
\frac{1}{a}\int_{-la/2}^{la/2}{ F\left[ 1 - \theta_{\rm c} (x') \right] dx'} \nonumber \\
& = & \frac{2D}{a^2} \theta^{*}( 1 - \theta^{\dagger}  ) + l F ( 1 - \bar{\theta_{\rm c}}) .
\end{eqnarray}
Analogously, loss and gain terms can be determined for the support region, so
that solutions for $\alpha_c$ and $\alpha_s$ are
\begin{eqnarray}
\alpha_{\rm c} = \frac{ \lambda_{\rm c} \left( r_{\rm s}-r_{\rm c}\right) \tanh{\left( \frac{da}{2\lambda_{\rm s}} \right) } } {\scriptstyle \lambda_{\rm c} \cosh{\left( \frac{la}{2\lambda_{\rm c}} \right)} \tanh{\left( \frac{da}{2\lambda_{\rm s}} \right)} + \left[ \lambda_{\rm s} + a\tanh{ \left( \frac{da} {2\lambda_{\rm s}} \right)} \right] \sinh{\left( \frac{la}{2\lambda_{\rm c}} \right) }} ,
\label{alphac}
\end{eqnarray}
and
\begin{equation}
\alpha_{\rm s} =\frac{ \lambda_{\rm s}(r_{\rm c}-r_{\rm s}) \tanh{\left( \frac{la}{2\lambda_{\rm c}} \right)} } {\scriptstyle \lambda_{\rm s}\cosh{\left( \frac{da}{2\lambda_{\rm s}} \right)} \tanh{\left( \frac{la}{2\lambda_{\rm c}} \right)} + \left[ \lambda_{\rm c} + a\tanh{\left( \frac{la} {2\lambda_{\rm c}} \right)} \right] \sinh{\left( \frac{da}{2\lambda_{\rm s}} \right)} } .
\label{alphas}
\end{equation}

The turnover frequency, which is the number of reactions per unit site and
unit time, is given by
\begin{equation}
TOF = \epsilon k_{\rm r} \bar{\theta_{\rm c}} .
\label{defdelta}
\end{equation}

% dar detalhes da simulacao
We simulated the discrete model, as defined in Sec. 2 (see also Fig. 1),
in order to check the accuracy of the analytic solution when $l$ and $d$
are not very large. Typical simulations consisted in 100 realizations of
the process in a lattice with $L=2^{17}$ sites and catalyst coverage near
$15\%$. At each step, an attempt to deposit a new reactant at a
randomly chosen site is done. Subsequently, the numbers of attempts to
move reactants, desorb them and perform reactions are chosen proportional to
the respective rates ($D$, $k_{\rm d}^{\rm s}$, $k_{\rm d}^{\rm c}$, $k_r$).
Simulations begin with an empty lattice and proceed up to a long time
after a steady state has been reached, where the coverages of both
regions are constant.

We observe that even for $l\sim 10$ the continuous approximation is
good. This
is illustrated in Fig. 2a and 2b, where we compare the analytic and numerical
results for the coverage distribution in the catalyst and in the support,
respectively, using $l=18$ and $d=100$. Slight deviations are only found near
the boundaries of those regions, but are always smaller than $5\%$ for small
$l$. The accuracy in the average coverages and in the TOF is usually
higher.

\begin{figure}[!h]
\begin{center}
\includegraphics[width=0.5\textwidth]{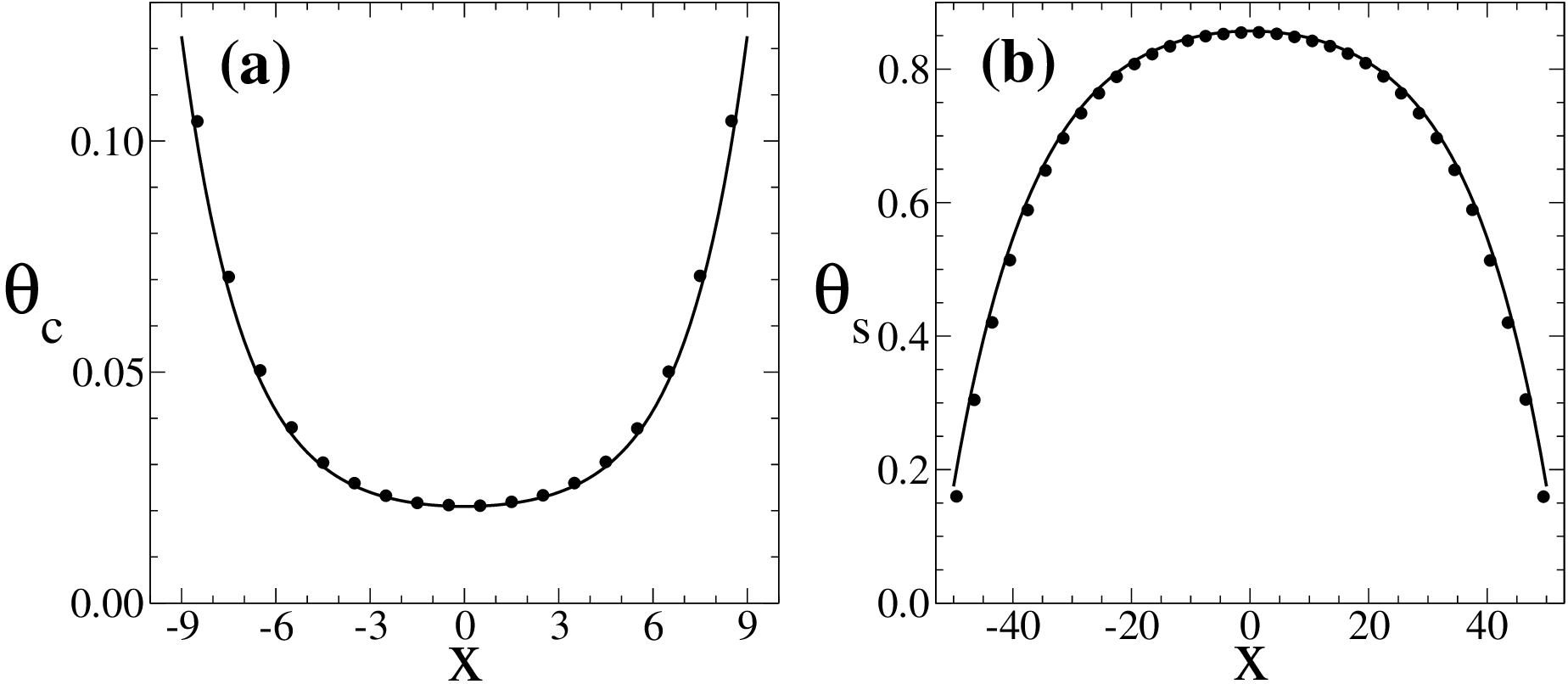}
\caption{\small Reactant coverage as a function of the position $x$ along (a) the catalyst and (b) the support, for $2D/\left( a^2F\right) =420$, $k_d^s/F=0.1$, 
$k_r/F=8$, $d=100$ and $l=18$. Circles are simulation results and the curves are analytical results.}
\label{fig2}
\end{center}
\end{figure}

Here the model was solved in the steady state, but it certainly can be extended
to other situations, for instance when a finite amount of the reactant flows to 
the catalyst surface and the TOF changes in time. Other
possibilities are the explicit incorporation of sticking probabilities and the
assumption of different diffusion coefficients in the particles and in the
support. The one-dimensional structure facilitates the solution, while
preserving essential ingredients such as the spatial heterogeneity.

% title change (ref. 2.5, 2.9)
\section{Results}

% explicacao sobre diffusion lengths (ref. 1.5)
In order to understand the role of the different rates from a scaling approach,
we consider the diffusion lengths in the catalyst particle and in the support.
If the corresponding region is large enough, the diffusion length measures the 
typical distance a reactant moves on it before reacting or desorbing. On the
other hand, if the length of that region is smaller than the diffusion length, 
then it is expected that the reactant reaches its border and can flow to a 
neighboring domain.

% mais detalhes sobre a derivacao destes valores a partir de RW properties
% (ref. 2.6)
In the catalyst, the average lifetime of an adsorbed species before reacting
or desorbing is $\tau_c\sim 1/\left( k_{r} + k_{d}^{c}\right)$. During this time, 
it executes random walks with diffusion coefficient $D$, thus the average 
distance it spans is of order ${\left( D\tau_c\right)}^{1/2}$. 
This is the so-called diffusion length,
\begin{equation}
L_{\rm c} =\sqrt{ \frac{ D }{ k_{r} + k_{d}^{c} } } ~.
\label{diflc}
\end{equation}
Analogously, a reactant in the support has a typical lifetime $1/k_{d}^{s}$,
thus the corresponding diffusion length is
\begin{equation}
L_{\rm s} = \sqrt{ \frac{ D }{ k_{d}^{s} } } ~.
\label{difls}
\end{equation}
These expressions can be easily generalized to the case of different diffusion
coefficients
in the particles and in the support, which means that interpretations based on
these quantities have a broader applicability.

We will consider cases where desorption in the catalytic region is very low,
i. e. $k_d^c\ll F,k_d^s,k_r,2D/a^2$. The values of all rates presented below are
given relatively to the incident flux rate $F$, thus they are all
dimensionless (setting $F=1~s^{-1}$ and the value of the parameter $a$, we would get
the other ones in SI units). The TOF is also expressed relatively to $F$, thus it
is limited to a maximum TOF/$F$ $=1$.

% explicacao sobre o uso do termo spillover a partir daqui
In order to facilitate the presentation of the results, hereafter we will use the
terms spillover and back spillover to denote a net flux of reactants by diffusion 
from the catalyst to the support and vice-versa, respectively. As
noted above, the broader use of these terms follows a trend of some recent works
\cite{zhdanovrev,costa,christou,galdikas,johansson,cwiklikSS}.

% subsection introduced
\subsection{Effects of reactant mobility and catalyst geometry}

First we distinguish the conditions where either spillover or back spillover is
dominant as the reactant mobility in the surface increases. In Figs. 3a and 3b
we show the normalized TOF as a function of $2D/a^2$ for several reaction
rates, respectively with fractions of support covered $\epsilon=0.05$ and
$\epsilon=0.15$, and the same particle size $l=75$.

\begin{figure}[!h]
\includegraphics[width=0.5\textwidth]{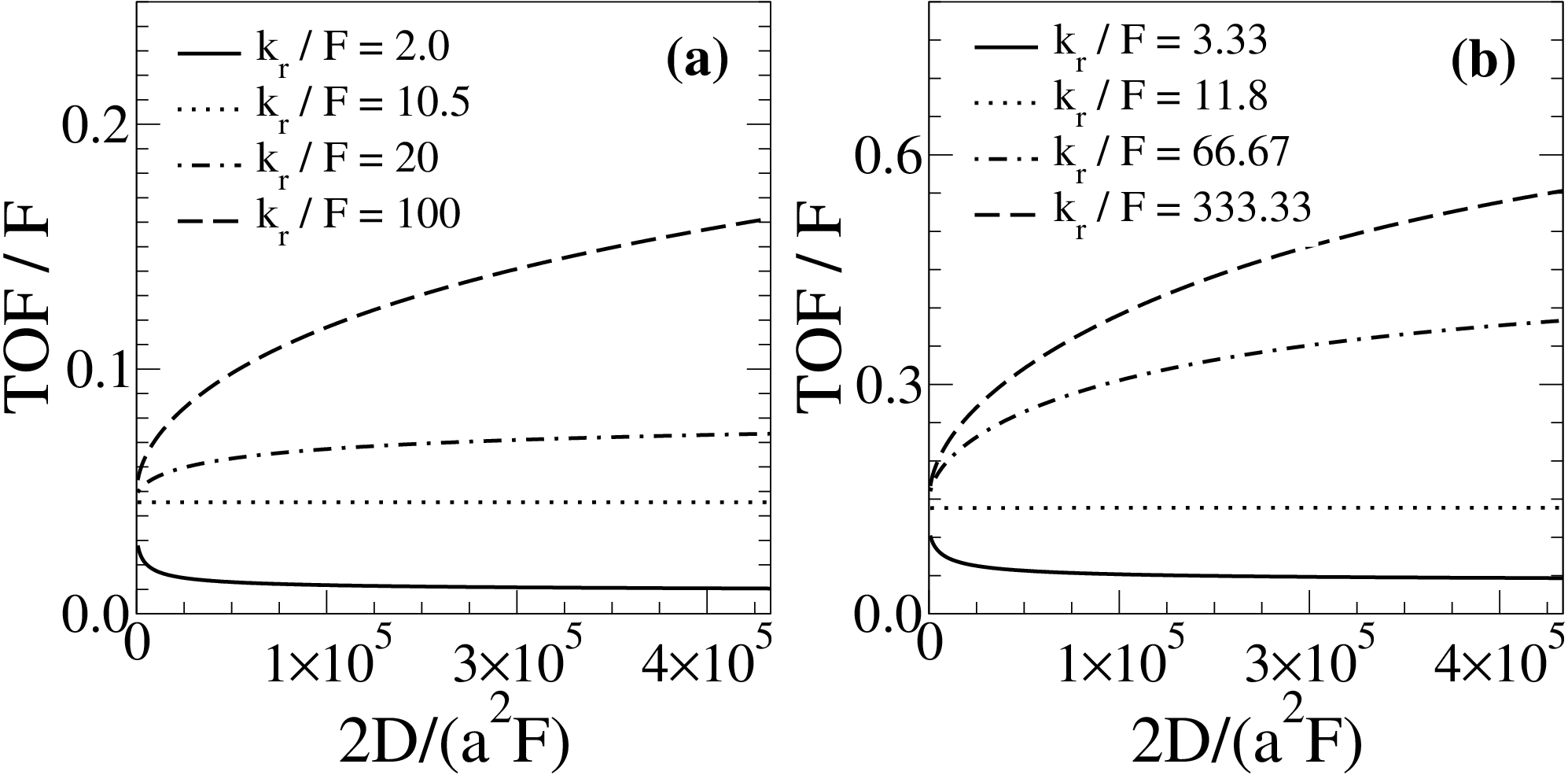}
\caption{\small Normalized turnover frequency as a function of  diffusion coefficient
for several reaction rates, with $k_d^c/F={10}^{-3}$, $k_d^s/F=10$ and $l=75$.
Fractions of support covered are (a) $\epsilon=0.05$ and (b) $\epsilon=0.15$.}
\label{fig3}
\end{figure}

% fornecer valores de TOF na fig. 3a para D=0 (ref. 1.13) - Thiago
% interpretar dependencia de Ls/Lc critico com epsilon (ref. 1.6)
Even without reactant diffusion ($D=0$), the TOF significantly varies
with $k_r$ because a large area of the catalyst may be poisoned for low reaction
rates; for instance, in Fig. 3a we have normalized TOF $= 0.0278$ and $0.0546$ for $k_r/F=0.1$ and
$k_r/F=5$, respectively. Indeed, excluded volume effects limit the adsorption process, 
which is known to be a relevant effect even in mean-field models \cite{lynggaard}.
The increase of diffusion coefficient improves the catalytic process for high
reaction rates, since back spillover effects are dominant as $D$ increases.
However, with low reaction rates, the opposite effect is observed: reactants
more easily leave the particles as $D$ increases, going to the support where
they rapidly desorb. In Fig. 3a, there is no change in the normalized TOF as $D$ increases
for $k_r/F=0.525$, which corresponds to $L_s/L_c\approx 0.23$. If other values of the rates and other values of $l$ are chosen, the same feature is observed for a different value of $k_r$, but with the same ratio $L_s/L_c$. On the other hand, in Fig. 3b ($\epsilon =0.15$), that feature is observed when $L_s/L_c\approx 0.42$. Thus,
the critical ratio $L_s/L_c$ which separates regimes of rapid and slow reactions depends only on the fraction of support covered $\epsilon$;
above (below) the critical ratio, back spillover (direct spillover) is dominant.
This result can be interpreted as follows: if the relative increase of diffusion length in the support is larger (smaller) than that in the catalyst, then more reactants flow towards the particles (support) and the TOF increases (decreases).

Now we consider the effects of particle size. We consider changes in
$l$ with fixed $\epsilon$, in order to simulate cases where a fixed amount of
catalytic material is deposited on the support, but islands of different sizes
are formed. Under these conditions, it is important to stress that the size of
the gaps between particles ($d$) also increases when $l$ increases (Eq. 
\ref{defrho}).

In Figs. 4a and 4b we show the normalized TOF as a function of the
particle size $l$ for two different fractions of the support covered and various
reaction rates. Again we observe regimes of high and low reaction
rates, corresponding to high and low ratios $L_s/L_c$. For large reaction
rates, decreasing $l$ is favorable, since $d$ also decreases ($\epsilon$ is
fixed)
and facilitates the back spillover.
On the other hand, for low reaction rates, the conversion is improved by
increasing the particle size, since the reactants spend longer times in
larger catalytic regions, which compensates the increased desorption in the
support. In other words, the loss due to decreasing back
spillover is compensated by a gain in decreasing spillover. The same ratios
$L_s/L_c$, which depend on the fraction of support covered $\epsilon$,
separate the two regimes.

\begin{figure}[!h]
\includegraphics[width=0.5\textwidth]{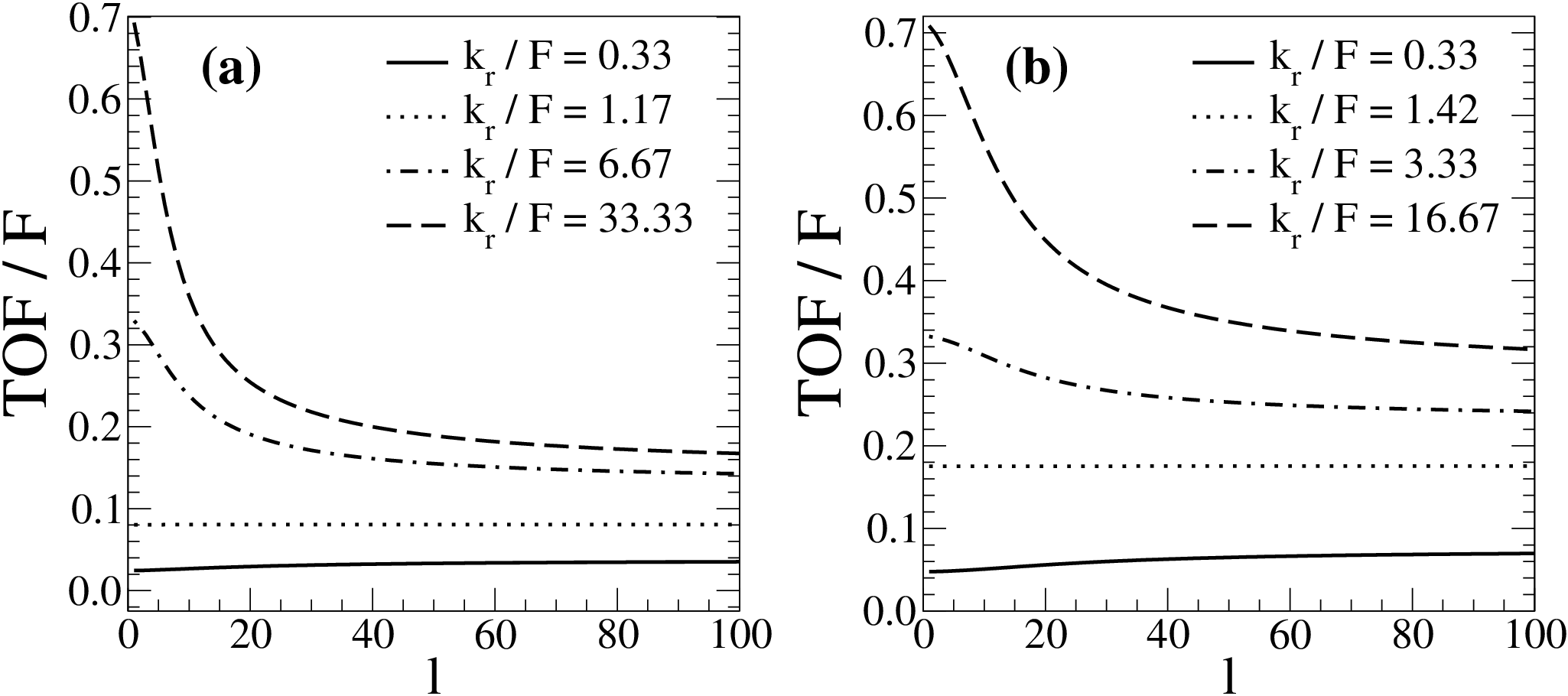}
\caption{\small Normalized turnover frequency as a function of catalyst particle size
for several reaction rates, using $k_d^c/F={10}^{-3}$, $k_d^s/F=1$ and $2D/\left( a^2 F\right) =500$. Fractions of support covered are (a) $\epsilon=0.15$ and (b) $\epsilon=0.3$.}
\label{fig4}
\end{figure}

% (ref 1.2)
% comentar resultados de Cwiklik em Chem Phys Lett 2007: sitios aleatorios sao
% melhores que distribuicao 1/2 porque o sistema esta' em um regime onde
% back spillover e' dominante e as reacoes sao rapidas, de modo que distancias
% pequenas entre particulas sao melhores.
% Note que o comprimento de difusao e' limitado no modelo deste artigo, pois
% pdiff<1 e o fluxo e' fixo. Esse deve ser o motivo pelo qual campo medio
% funciona bem. Provavelmente sera' dificil fazer campo medio com difusao
% rapida, pois a vizinhanca de um reagente e'  muito larga.

These nontrivial effects of the catalyst geometry may be helpful for catalyst
design if one is interested in taking
advantage of back spillover, preferrably in the case of rapid reactions, or in
reducing the effect of spillover in the case of slow reactions. At this point,
it is important to notice that Figs. 3 and 4 show that the change in one
model parameter may increase the TOF by a factor near $3$, which
is a remarkable change in the efficiency of the catalytic process.

% comentar resultados de Cwiklik em Chem Phys Lett 2007 (ref 1.2)
As far as we know, this
is the first work that analyzes the conditions for changing the direction of
the net reactant flux in the particle-support interface.
The interpretation of results based on diffusion lengths was formerly
proposed in Refs. \protect\cite{mcleod,zhdanov1997} for models of $CO$ oxidation
in oxide supports. However, both studies focused on the regime where back spillover
is dominant, for instance by assuming infinite diffusion
lengths of some species adsorbed in the particles. That regime was also
considered in Ref. \protect\cite{cwiklikCPL} with a mean-field approach that
accounts for the different neighborhood of the catalytic sites (in an approximate
form). In this case, the approximation is successful because the diffusion
lengths are small.

% section -> subsection (ref. 2.9)
\subsection{Temperature effects}

% escrever nu_i e E_i para designar amplitudes e energias em geral (ref. 1.14)
The diffusion coefficient and the reaction and desorption rates are expected to
have
Arrhenius forms as follows:
\begin{equation}
D = \frac{a^2}{2} \nu_{\rm D} \exp{ \left( -\frac{ E_{\rm D} }{ k_B T }
\right) },
\label{Darr}
\end{equation}

\begin{equation}
k_{\rm d}^{\rm s} = \nu_{\rm d} \textrm{exp} \left( -\frac{ E_{\rm d}}{ k_B T }
\right) ,
\label{kdsarr}
\end{equation}
and
\begin{equation}
k_{\rm r} = \nu_{\rm r} \exp{ \left( -\frac{ E_{\rm r} }{ k_B T }
\right) } .
\label{krarr}
\end{equation}
Here, $\nu_i$ is a frequency and $E_i$ is an activation energy ($i=D,d,r$).
Assuming that
the activation energy for desorption in the catalyst particles is much larger
than the other activation energies, we use $k_d^c/F={10}^{-3}$, which is
negligible
compared to the other rates in the relevant temperature ranges. For the other
activation energies, we always assume that $E_D<E_d$ \cite{gomer}.

Other reasonable assumptions on the amplitudes of the Arrhenius
rates facilitate the analysis of the effects of different ranges of
energy barriers. We will consider $\nu_D=\nu_r=2\times {10}^{12}~
s^{-1}$, $a=5~\mathring A$,
and work with a range of ratios $\nu_d/\nu_r$ from $10$ to $1000$. These
relations
are reasonable for Langmuir-Hinshelwood (LH) reactions and for $CO$ adsorption
in
oxides
\cite{zhdanovrev,zhdanov1997,gomer}, respectively, but we emphasize that these
systems are only rough guides to choose working parameters, and may not be
viewed as
prospective applications - indeed, the unimolecular reaction of our model is
representative of ER mechanism.
In the following, we also consider fraction of support covered $\epsilon=0.15$
and $F=1~ s^{-1}$ (thus the calculated rates are again ratios to $F$). 

Qualitatively, we expect that each microscopic process will be significantly
activated when its rate exceeds the external flux $F$. However,
excluded volume effects lead to surface poisoning when reactions are not
frequent (low temperatures), so that other processes can affect the turnover
frequency only after reactions are activated. 

% subsubsection introduced (ref. 2.10, 2.11)
\subsubsection{The cases $E_r>E_d$ and $E_r<E_D$}

% explicacao do limite TOF=0.15 (ref. 1.7)
% exemplo de reacao lenta (ref 5)
First we consider the simplest case where reactions are very difficult compared
to the other activated processes, i. e. $E_r>E_d$. An example of slow reaction
is the hydrogenation of $CO$ on $Pt$ \cite{mao}, when compared to the spillover
to the $TiO_2$ support and the formation of $CH_3O$ there.

Fig. 5a shows the typical
evolution of the TOF with temperature for different particle
sizes. When the reactions become more frequent ($k_r\sim F$, i. e. $T\sim
270~K$),
desorption is already activated and diffusion is very fast. Thus spillover and
subsequent desorption of reactants does not allow the increase of the turnover
frequency. Instead, the TOF begins to increase only when $k_r\sim k_d$
($T\sim 550 ~K$). At higher temperatures, the negative contribution of spillover 
is more important when the particles are small, so that a slow increase of the
TOF is observed. On the other hand, this negative contribution is reduced for
large particles, and the maximum TOF is rapidly attained by increasing the
temperature. The value TOF$/F\approx 0.15$ corresponds to the fraction of the surface covered by the catalyst, which is expected at high temperatures
because only species adsorbed on the catalyst react.

\begin{figure}[!h]
\includegraphics[width=0.5\textwidth]{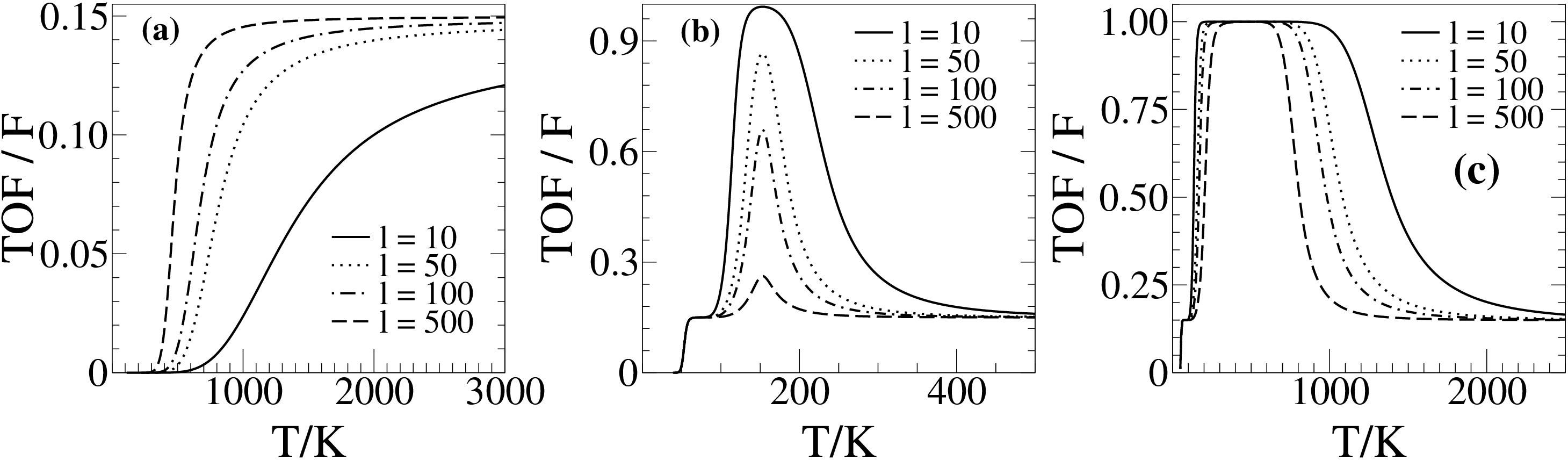}
\caption{\small Normalized turnover frequency as a function of temperature for several particle sizes, with: (a) $E_D=5~ kcal/mol$, $E_r=15~ kcal/mol$ and $E_d=10~ kcal/mol$; (b) $E_D=5~ kcal/mol$, $E_r=3~ kcal/mol$ and $E_d=10~ kcal/mol$; (c) $E_D=6~ kcal/mol$, $E_r=3~ kcal/mol$ and $E_d=35~ kcal/mol$. In all cases, $\nu_d/\nu_r=100$.}
\label{fig5}
\end{figure}

% example of rapid reaction (ref 5)
Next we consider the case where reactions are easily excited, i. e. $E_r<E_D$.
An example of rapid reaction with ER mechanism is $CO$ oxidation on $Pd/CeO_2$
\cite{costa}, where reaction rates are nearly $100$ times larger than
back spillover rates of oxygen.

Figs. 5b and 5c illustrate the case $E_r<E_D$ with small and large desorption
energies, respectively.

In Fig. 5b, the TOF increases towards a
plateau in TOF$/F\approx 0.15$  at $T\sim 50-60 ~K$), where $k_r\sim F$.
At $T\sim 100~ K$, when diffusion is activated, back spillover leads to a second
jump in the TOF, which is highly dependent on the particle size.
For smaller sizes (small gaps between particles), the migration of reactants
from the support to the catalyst is easy even for low $D$, thus large
conversion rates are rapidly obtained. The temperature of maximal TOF is
attained when desorption begins to play a significant role ($k_d^s\sim F$),
independently of particle size. In the case of large
gaps between the particles, this temperature is still low for back spillover to
be efficient, thus only a small peak appears in the TOF plot. For further
temperature increase, $l$-dependent results are again obtained. The diffusion
length in the support is
$L_s= \sqrt{\frac{\nu_D a^2}{\nu_d}}\exp{\left[ \left( E_d-E_D\right) / 2k_B
T\right]}$,
which decreases with increasing temperature, and the
beneficial effect of back spillover ceases when $L_s\sim l$; this condition is
satistified at higher temperatures for smaller $l$, which explains the
slower decay of the TOF in this case. For these reasons, the peak in the TOF
is high and broad for small $l$, low and narrow for large $l$.

In Fig. 5c, the main features of Fig. 5b are present. However, since $E_d$ is
much larger than the other activation energies, the maximal effect of
back spillover (TOF/F $\approx 1$) is observed in a wider temperature range and
for all particle sizes shown there.
Thus, Fig. 5b illustrates typical conditions in which
particle size effects are clearer: desorption in the support has higher
activation energy than diffusion and reactions, but activation of a process
begins while the other processes are not fully activated, so that the diffusion
length $L_s$ cannot attain large values before desorption is
activated.

% subsubsection introduced (ref. 2.12)
\subsubsection{The case of intermediate reaction energies}

%exemplo de reacao intermediaria (ref 5)
Now we consider cases of intermediate activation energies for reaction, i. e.
$E_D<E_r<E_d$. $CO$ oxidation provides several examples with such relation
between activation energies; however, it is important to stress the difference
in the usual reaction mechanism (LH instead of ER). An example is $CO$
oxidation on $Pt/CeO_2$
\cite{johansson}: the energy of diffusion of $O$ on the support is
$18 kcal/mol$, while reaction energy is $27 kcal/mol$ and desorption in the
support is $60 kcal/mol$ (diffusion on the catalyst is assumed to be fast for
the application of a mean-field model in Ref. \protect\cite{johansson}).

In Fig. 6a we show the normalized TOF as a function of temperature for our
model with three different particle sizes, using a set of activation energies
previously suggested for $CO$ oxidation in $Pt(111)$ and $\nu_d=100\nu_r$
\cite{zhdanovrev}. Fig. 6b shows the
evolution of the average coverage (particle plus support).

\begin{figure}[!h]
\includegraphics[width=0.5\textwidth]{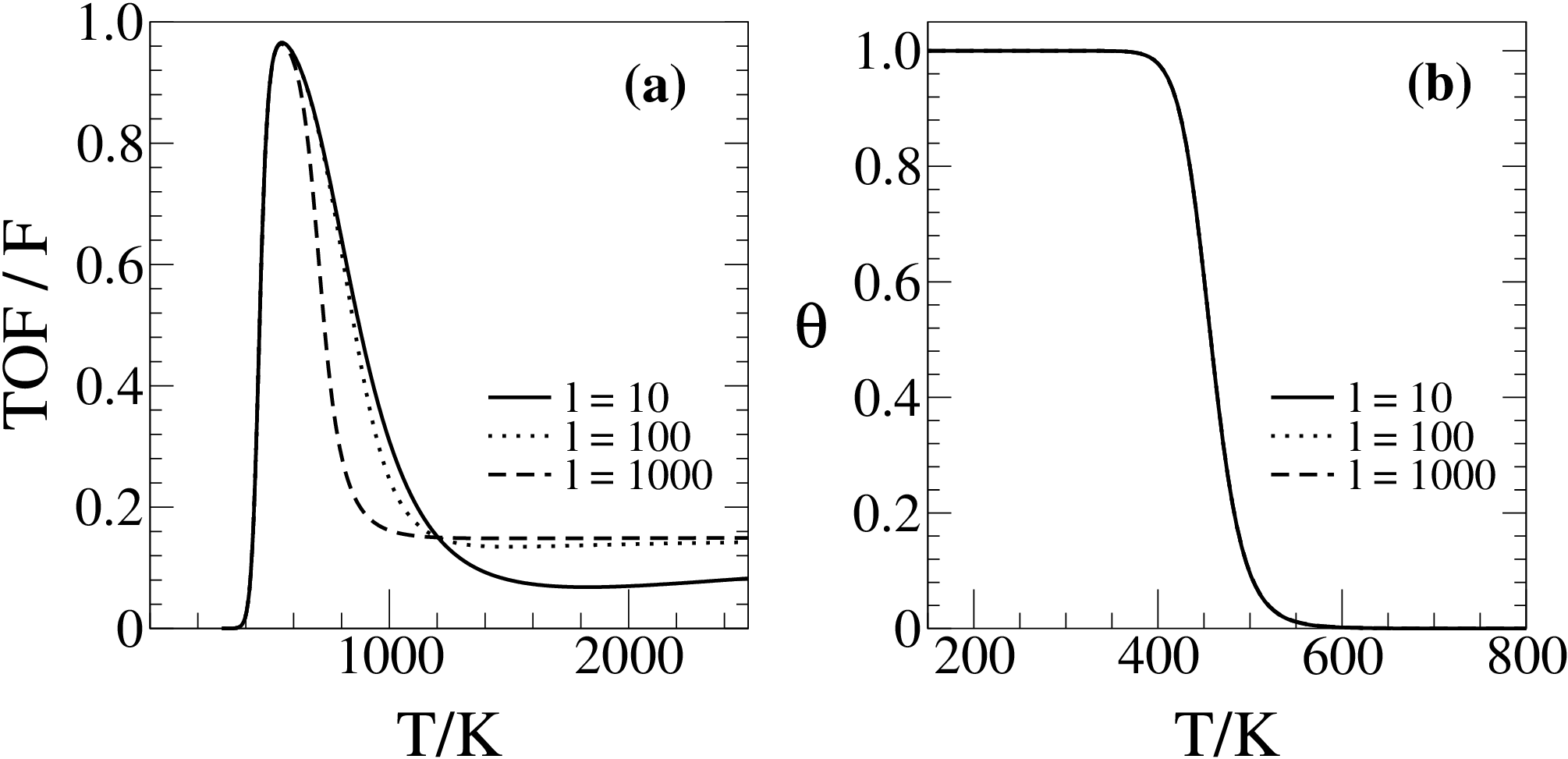}
\caption{\small (a) Normalized turnover frequency as a function of temperature for several particle sizes, with $E_D=6~ kcal/mol$, $E_r=24~ kcal/mol$ $E_d=35~ kcal/mol$, and $\nu_d/\nu_r=100$. (b) Temperature dependence of the corresponding average reactant coverage.}
\label{fig6}
\end{figure}

At low temperatures, all the surface is poisoned, and reactions are slow.
Increasing the temperature to $T\sim 400~ K$ ($k_r\sim F$), the turnover
frequency
increases, thus
vacancies are created in the particles and in the support, which facilitates the
back
spillover effects.
Note that there is no significant effect of particle size when a
large temperature range is scanned, since the diffusion lengths $L_s$ and $L_c$
are very large  (diffusion is highly activated in much lower temperatures).
While the TOF increases with temperature, the coverage rapidly
decreases towards zero because diffusion
and reaction are rapid compared to the external flux.
Further temperature increase leads to a maximum in the turnover frequency,
again when desorption in the support is activated ($k_d^s\sim F$). However,
in the right side of the peaks of Fig. 6a, we observe a size dependence
because back spillover ceases only when $L_s\sim l$. At high
temperatures, a size dependence is also noticeable: small particles
lose more reactants by spillover than the large ones, and those reactants easily
desorb in the support, which leads to smaller TOF.

% observacao de que E_D e' menor que E_r, mas e' proximo (ref. 1.15)
As explained above, the most remarkable effects of particle size are observed
in cases where the activation energies are close to each other. This is also
illustrated in Figs. 7a and 7b for $E_D<E_r<E_d$, but $E_D$ close to $E_r$,
with different ratios between $\nu_r$ and
$\nu_d$. In this case, when reactions become more frequent and leave room
for new incoming reactants ($k_r\sim F$), the surface mobility is still low
(i. e. $2D/a^2$ is not much larger than $F$). Thus, back spillover is
significant
only for small particles (and small
vacancies between them). However, if the spacing between
particles is large, then slow diffusion is not
able to bring all reactants adsorbed in the support to the catalytic region.
Figs. 7a and 7b also show that these effects appear in a wider temperature
range when the amplitudes of the Boltzmann factors are closer to each other,
since activation of desorption occurs in higher temperatures.

\begin{figure}[!h]
\includegraphics[width=0.5\textwidth]{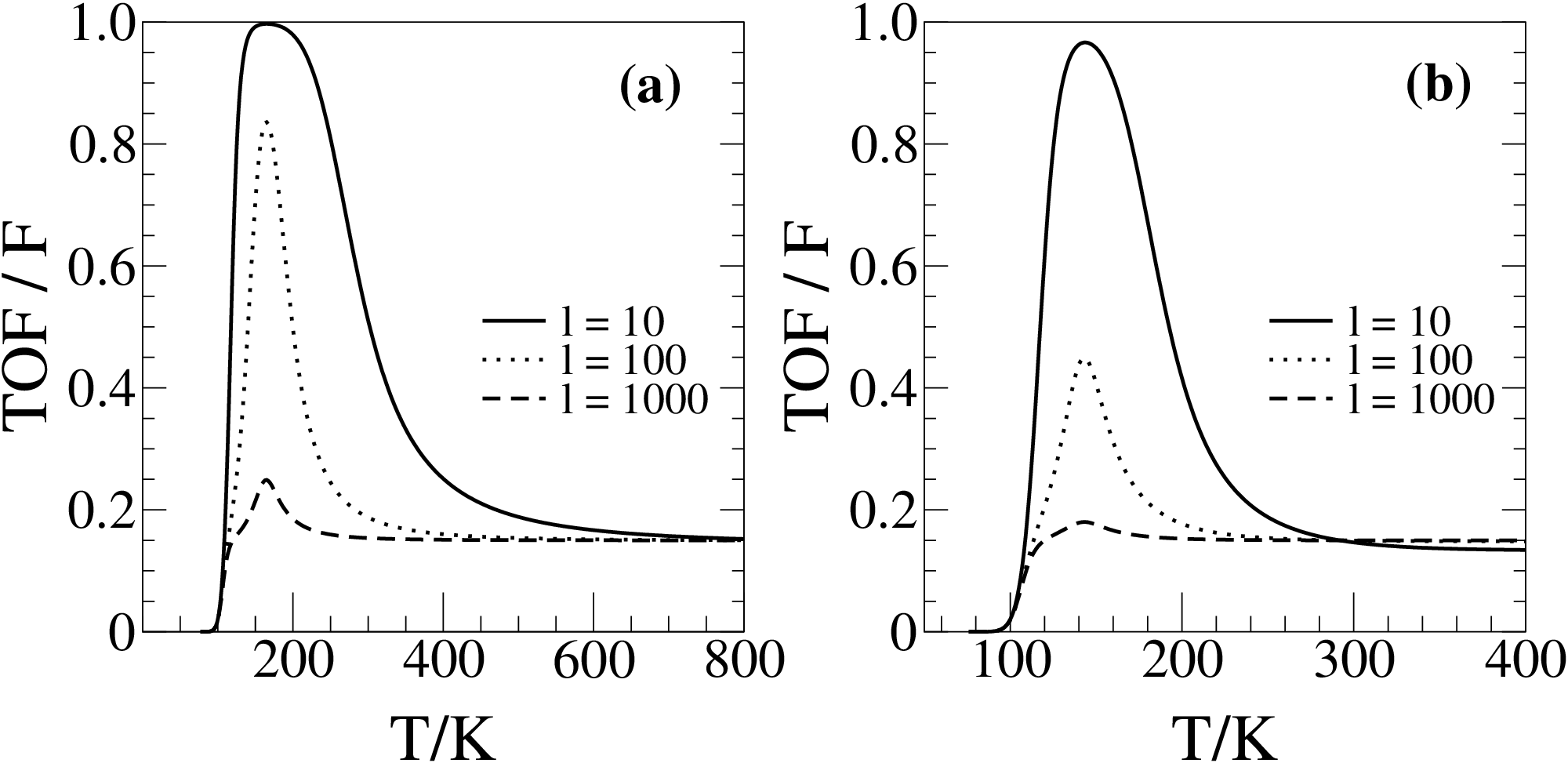}
\caption{\small Normalized turnover frequency as a function of temperature for several particle sizes, with $E_D=5 ~kcal/mol$, $E_r=6~ kcal/mol$ $E_d=10~ kcal/mol$: (a)$\nu_d/\nu_r=10$; (b) $\nu_d/\nu_r=1000$.}
\label{fig7}
\end{figure}
 
\section{Conclusion}

We proposed a model for a unimolecular reaction in a supported catalyst with
tunable particle size and fraction of support covered. It represents important
physico-chemical processes, such as diffusion, desorption and the external
flux of reactants. We analyzed the
effects of spillover and back spillover on the turnover frequency under
different conditions.

First we considered the isolated effects of enhancing
surface diffusion (coefficient $D$) and increasing catalyst particle size $l$.
For rapid reactions, the increase of any of those quantities favors back
spillover
and, consequently, increases the turnover frequency. This regime is separated
from that dominated by spillover by a critical value of diffusion length
in the catalyst and in the support, which depends only on the fraction of
the support surface covered by the catalyst. In the spillover-dominated
regime (slow reactions), increasing $D$ or $l$ slows down the conversion of
the reactants.

% small change (ref. 2.14)
Subsequently, we considered temperature effects by assuming Arrhenius
dependence of all physico-chemical rates and reasonable values for the
amplitudes in those relations. With activation energy for desorption ($E_d$)
in the support larger than that for diffusion ($E_D$), a peak in the turnover
frequency as a function of temperature is observed for small and intermediate
values of reaction activation energy $E_r$, i. e. for cases where $E_r<E_d$.
Significant particle size dependence in the peaks is usually observed when
the gaps between those energies are small, so that activation of one process
occurs while the other ones are not fully activated yet, and the corresponding
diffusion lengths rapidly vary with temperature. The right side of
those peaks show size-dependence under more general conditions.
For fixed amount of catalytic
material deposited on the support, small particle sizes (with small distance
between them) allow the turnover frequency to attain high peak values due to the
beneficial effect of back spillover, while large particles provide low
enhancements of catalytic activity. Finally, in the case of slow
reactions ($E_r>E_d$), the TOF monotonically increases with
temperature, and large particle sizes are more efficient to avoid the
negative effects of direct spillover.

The aim of the present work is not the quantitative description of a particular
catalytic process, but to discuss the interplay of various physico-chemical
mechanisms and conditions in systems where spillover is present. However, it
may be useful in the interpretation of some experimental results and motivate
the proposal of extended models for their quantitative description.

%mudancas
Here we mention two recent works as examples where the scaling ideas
developed above may be relevant. However, we emphasize the fact that our model
cannot be directly applied to these systems, thus our results only suggest
general guidelines to understand their qualitative behavior. In the first
example,
Piccolo and Henry \cite{piccolo} studied the oxidation of $CO$ by $NO$ on
$Pd/MgO$
and observed a peak in the TOF as a function of the temperature.
They found a remarkable increase in the peak height as the particle size was
decreased
(even being accompanied by a decrease in the $Pd$ coverage), but tiny shifts in
the temperature of the maximum. These results resemble those in Figs. 6a, 7a and
7b. While the size-independent curve shape suggests weak size-dependence of
activation
energies, the peak height increase suggests that diffusion lengths of reactants
were
not very large when reactions were activated, so that back spillover (which
increases
TOF for any type of reaction) was facilitated with small particles. Thus,
differences in activation energies of diffusion, reaction and desorption are
probably
small, with $E_d$ being larger than the other ones. The second example is
a study of $CO$ oxidation on $Au$ on active supports (i. e.
those which adsorb and supply oxygen to the reaction), which
shows a different trend \cite{schubert}: under
certain conditions, the TOF does not depend on catalyst particle size.
Although that reaction involves two species and is probably of LH
type, the experimental result suggests that diffusion was highly activated at
the
working temperature. In this situation, similarly to our model, diffusion
lengths are always
very large and the contribution of back spillover does not depend on particle
size.

% changes
Recalling the results of models for ethene and acetylene hydrogenation which
incorporate
spillover effects is also interesting at this point
\cite{mcleodCERD2004,mcleodCES2004}.
In both systems, the hydrocarbon blocks catalytic sites at high pressures, which
leads
to a discontinuity of the TOF when the external reactant flux towards a
catalytic site
is of the same order of the microscopic reaction rate. In ethene hydrogenation,
this
effect is shown to be more intense with small particles \cite{mcleodCERD2004},
leading
to finite-size effects. However, diffusion is
weakly activated at the working temperature when compared to the other processes
(the
amplitudes of Boltzmann factors associated to different processes are very
different
in that case), thus there is no significant effect of diffusion in the TOF.
This reinforces the conclusion that the scaling approach proposed here can be
extended
to interpret more complex reaction models. This is very
important in cases where analytic solutions are not feasible
because it may help choosing the conditions to perform simulation work.

\vskip 0.5cm

{\par\noindent\bf Acknowledgments.} The authors thank Prof. Robert Hamers for a
critical
reading of the manuscript and acknowledge support from CNPq and
Faperj (Brazilian agencies) to their simulation laboratory at UFF, Brazil.
TGM acknowledges a grant from CAPES (Brazil) and FDAAR acknowledges support from
CNPq for his visit to UW - Madison.


\begin{thebibliography}{99}

\bibitem{broadbelt2000}
L. J. Broadbelt, R. Q. Snurr, Appl. Catal. A: Gen. 200 (2000) 23.

\bibitem{zhdanovrev}
V. P. Zhdanov, B. Kasemo, Surf. Sci. Rep. 39 (2000) 25.

\bibitem{lynggaard}
H. Lynggaard, A. Andreasen, C. Stegelmann, P. Stoltze, Prog. Surf. Sci. 77
(2004) 71.

\bibitem{murzin}
D. Y. Murzin, Ind. Eng. Chem. Res. 44 (2005) 1688.

\bibitem{piccolo}
R. Piccolo, C. R. Henry, Appl. Surf. Sci. 162 (2000) 670.

\bibitem{chen}
H. Chen, H. Yang, Y. Briker, C. Fairbridge, O. Omotoso, L. Ding, Y. Zheng,
Z. Ring, Catal. Today 125 (2007) 256.

\bibitem{polychrono}
K. Polychronopoulou, A. M. Efstathiou, Catal. Today 116 (2006) 341. 

\bibitem{kecskemeti}
A. Kecskem\'eti, T. B\'ans\'agi, F. Solymosi, Catal. Lett. 116 (2007) 101.


\bibitem{laurin}
M. Laurin, V. Joh\'anek, A. W. Grant, B. Kasemo, J. Libuda, H.-J. Freund, J.
Chem. Phys. 123 (2005) 054701.

\bibitem{odier}
E. Odier, Y. Schuurman, C. Mirodatos, Catal. Today 127 (2007) 230.

\bibitem{marques}
R. Marques, S. Capela, S. DaCosta, F. Delacroix, G. Djega-Mariadassou, P. Da
Costa, Catal. Comm. 9 (2008) 1704.

\bibitem{bowker}
M. Bowker and E. Fourr\'e, Appl. Surf. Sci. 254 (2008) 4225.

\bibitem{mao}
T. F. Mao and J. L. Falconer, J. Catal. 123 (1990) 443.

\bibitem{lueking}
A. D. Lueking, R. T. Yang, Appl. Catal. A: General 265 (2004) 259.

\bibitem{dutta}
G. Dutta, U. V. Waghmare, T. Baidya, M. S. Hegde, Chem. Mat. 19 (2007) 6430.

\bibitem{jain}
P. Jain, D. A. Fonseca, E. Schaible, A. D. Lueking, J. Phys. Chem. C 111 (2007)
1788.

%\bibitem{lachawiec}
%A. J. Lachawiec, Jr., G. Qi, R. T. Yang, Langmuir 21 (2005) 11418.

%\bibitem{panayotov}
%D. A. Panayotov, J. T. Yates, Jr., J. Phys. Chem. C 111 (2007) 2959.

\bibitem{henryrev}
C. R. Henry, Surf. Sci. Rep. 31 (1998) 231.

\bibitem{conner}
W. C. Conner and J. L. Falconer, Chem. Rev. 95 (1995) 759.

\bibitem{libuda}
J. Libuda, H.-J. Freund, Surf. Sci. Rep. 57 (2005) 157.

\bibitem{hoffman}
J. Hoffmann, I. Meusel, J. Hartmann, J. Libuda, H.-J. Freund, J. Catal. 204
(2001) 378.

\bibitem{costa}
C. N. Costa, S. Y. Christou, G. Georgiou, A. M. Efstathiou, J. Catal. 219 (2003)
259.

\bibitem{christou}
S. Y. Christou, A. M. Efstathiou, Top. Catal. 42-43 (2007) 351.

\bibitem{galdikas}
A. Galdikas, D. Duprez, C. Descorme, Appl. Surf. Sci. 236 (2004) 342.

\bibitem{dooling1999}
D. J. Dooling, J. E. Rekoske, L. J. Broadbelt, Langmuir 15 (1999) 5846.

\bibitem{cwiklikCPL}
L. Cwiklik, Chem. Phys. Lett. 449 (2007) 304.

\bibitem{henryJCP}
C. R. Henry, C. Chapon, C. Duriez, J. Chem. Phys. 95 (1991) 700.

\bibitem{mcleod}
A. S. McLeod, Catal. Today 53 (1999) 289.

\bibitem{zhdanov1997}
V. P. Zhdanov, B. Kasemo, J. Catal. 170 (1997) 377.

%\bibitem{volkening}
%S. Volkening, J. Wintterlin, J. Chem. Phys. 114 (2001) 6382.

\bibitem{johansson}
S. Johansson, L. \"Osterlund, and B. Kasemo, J. Catal. 201 (2001) 275.

\bibitem{mcleodCERD2004}
A. S. McLeod, Chem. Eng. Res. Design 82 (2004) 945.

\bibitem{mcleodCES2004}
A. S. McLeod and R. Blackwell, Chem. Eng. Sci. 59 (2004) 4715.

\bibitem{cwiklikSS}
L. Cwiklik, B. Jagoda-Cwiklik, M. Frankowicz, Surf. Sci. 572 (2004) 318.

\bibitem{cwiklikASS}
L. Cwiklik, B. Jagoda-Cwiklik, M. Frankowicz, Appl. Surf. Sci. 252 (2005) 778.

\bibitem{oshaninJCP}
G. Oshanin, A. Blumen, J. Chem. Phys. 108 (1998) 1140.

\bibitem{benichou}
O. B\'enichou, M. Coppey, M. Moreau, G. Oshanin,	J. Chem. Phys. 123 (2005)
194506.

\bibitem{oshaninPRL}
G. Oshanin, M. N. Popescu, S. Dietrich, Phys. Rev. Lett. 93 (2004) 020602.

\bibitem{albano1}
E. V. Albano, J. Chem. Phys. 94 (1991) 1499.

\bibitem{albano2}
E. V. Albano, Phys. Rev. E 48 (1993) 913.

%\bibitem{mcleodJCICS2000}
%A. S. McLeod and L. F. Gladden, J. Chem. Inf. Comput. Sci. 40 (2000) 981.


\bibitem{jansenPRL}
A. P. J. Jansen, C. G. M. Hermse, Phys. Rev. Lett. 83 (1999) 3673.

\bibitem{rieger}
M. Rieger, J. Rogal, K. Reuter, Phys. Rev. Lett. 100 (2008) 016105.

\bibitem{oshaninJSP}
G. Oshanin, O. B\'enichou, A. Blumen, J. Stat. Phys. 112 (2003) 541.

\bibitem{incubation}
F. D. A. A. Reis, J. Stafiej, J.-P. Badiali, J. Phys. Chem. B 110 (2006) 17554.

\bibitem{somorjai}
G. A. Somorjai, R. L. York, D. Butcherab, J. Y. Parka, Phys. Chem. Chem. Phys. 9
(2007) 3500.

\bibitem{schuth}
F. Sch\"uth, Annu. Rev. Mater. Res. 35 (2005) 209.

\bibitem{freund2002}
H.-J. Freund, Surf. Sci. 500 (2002) 271.

\bibitem{gomer}
R. Gomer, Rep. Prog. Phys. 53 (1990) 917.

\bibitem{schubert}
M. M. Schubert, S. Hackenberg, A. C. van Veen, M. Muhler, V. Plzak, R. J. Behm, 
J. Catal. 197 (2001) 113.


\end{thebibliography}
\end{document}